# Resonant interatomic Auger transition in chalcopyrite CuInSe2

Vladimir I. Grebennikov [*,1,2] and Tatyana V. Kuznetsova [1,2]

[1]Mikheev Institute of Metal Physics, Ural Branch, Russian Academy of Sciences, Ekaterinburg, 620108 Russia
[2]Ural Federal University named after the first President of Russia B.N. Yeltsin, Ekaterinburg, 620002 Russia
*e-mail: *vgrebennikov@list.ru*

The interatomic Auger transitions in compounds containing atomic components with core levels close in energy are studied theoretically and experimentally. The Coulomb transitions of a hole between such levels lead to a resonant enhancement of the Auger spectra (with respect to the energy difference between the levels). Interatomic Auger transitions involving high lying levels are formed by shaking up electrons due to the dynamic field of photo holes produced during the transition. These effects were observed experimentally in XPS and Auger spectra of CuInSe2 type materials.

## 1. Introduction

The study of interatomic Auger transitions is of fundamental importance for understanding the evolution of excited states in matter. It was stated[1-2] that the photoemission associated with a certain electronic level of the given atom "A" can be significantly changed in intensity as the photon energy passes through the absorption edge of the core level of the neighboring atom "B". This effect was called a multiatomic resonant photoemission (MARPE). Its macroscopic analogue can be the so-called resonant X-ray optical dielectric model, which considers the change in the complex dielectric constant of the substance as it passes through the corresponding resonance.[3-5] The variation of the Auger process is the so-called Interatomic Coulombic Decay (ICD) in which the excitation of valence electrons of one atom leads to the ionization of the valence shell of the second atom. ICD is considered as a new source of low energy electrons in slow ion-dimer collisions.[6] Inelastic scattering of Auger electrons and interatomic Coulombic decay are suggested as the mechanisms populating and depopulating, respectively, the excited states in nanoplasma which are formed when an X-ray free-electron laser pulse interacts with nanometer-scale matter, for example with a cluster of xenon atoms.[7] ICD can serve as the main mechanism for rapid deexcitation of hollow atoms arising from the neutralization of a highly charged ion in solids.[8] Recently, the single-photon laser-enabled Auger decay in Ne atoms decay was experimentally detected.[9] Most of the investigations was aimed at the study of atomic transitions involving valence electrons. In solid state physics, the objects were oxides, for example, $TiO_2$ [10] or sodium halides,[11] whose valence bands are formed mainly by metalloid states. But even in such compounds, it turned out to be difficult to separate, for example, $TiL_{23}M_{23}V(O)$ and $TiL_{23}M_{23}V(Ti)$ Auger transitions due to the close energies of the final states. As a result, the question of the direct observation of interatomic Auger transitions in the X-ray energy range remains open. Since our experiments were performed on photovoltaic materials, we noticed that Auger processes play an important role in photoelectric converters since they determine energy losses and limit the efficiency of their operation[12], of course in the optical and not in the X-ray range, which will be considered in this article.

## 2. Auger Process in Compound with Element Levels Close by Energy

Let us consider the scheme (**Figure 1a**) of the interatomic Auger transition using the CuInSe2 compound as an example. The photoionized Cu2p hole (binding energy 933 eV, labeled as c) is filled by an electron from the overlying Cu3p state (75 eV, A1) due to the Coulomb interaction, and the energy released in this process is spent on the ejection of an electron located on the neighboring indium atom In4d (17 eV, B2) into free state f. The detector measures the intensity of the outgoing electrons versus their kinetic energy or Auger spectrum of CuL3M2,3InN4,5. Usually the intensity of the interatomic transition is vanishingly small in comparison with the intra-atomic transition, say CuL3M2,3V (V denotes a valence state), since the distance determining the energy of the Coulomb interaction $e^2/|\mathbf{r}_1 - \mathbf{r}_2|$ between electrons inside the atom is much smaller than the interatomic distance.

We will show that the intensity of the interatomic Auger transition substantially increases if the neighboring element has a core level close in energy to the level of the central atom. In our example, this is an In4p level ($B_1$, with a binding energy of 73.5 eV, which is shifted by only 1.5 eV from the energy of the Cu3p level. In this case, the possibility of the real (in the final state) transition of a hole from one atom to another (Figure 1b) increases, as well as the probability of virtual hole transitions between levels (Figure 1c).

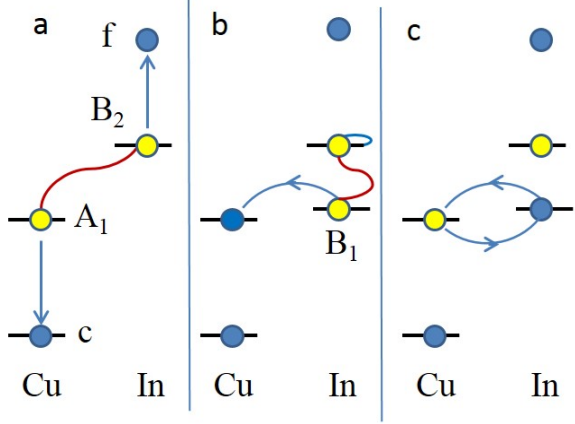

**Figure1.** Diagrams of the interatomic Auger process Cu2pCu3pIn4d in the CuInSe2 compound (a) considering the real (b) and virtual (c) transitions between the Cu3p ($A_1$) and In4p ($B_2$) levels. The In3p level is a labeled $B_1$ used in the equations for a quick identification.

The transition rate from the initial state i to the final state j (generally damped) is determined by the equation

$$I_{ji} = \frac{1}{\hbar\pi} \operatorname{Im} G_{jj}(E_i) |A_{ji}(E_i)|^2 ;$$
$$A_{ji}(E) = (V + VG(E)V)_{ji} \quad (1)$$

Here

$$G(E) = (E - H - V - i\gamma)^{-1} \quad (2)$$

is the total Green's function with the Coulomb interaction $V$ causing the Auger transitions between the eigenstates i and j of the basic Hamiltonian H. The value $A_{ji}(E)$ defines the transition amplitude, and the imaginary part $\operatorname{Im} G_{jj}(E_i)$ describes the energy conservation law with allowance for the damping of the initial and final states; $\gamma$ is level half-width. The last is assumed to be the same in the final and intermediate states for simplicity.

In the lowest order, the amplitude of the Auger transition is equal to the matrix element of the Coulomb interaction between the initial and final states

$$A_{ji} = V_{ji} = <A_1, B_2 | V | c, f> . \quad (3)$$

With reference to the diagram in Figure 1, the state c is the hole at the Cu2p level, f is the electron with energy $E_f$ counted by the detector, state $A_1$ is the Cu3p hole, and $B_2$ is the hole at the In4d level. The amplitude (Figure 1a) of the Auger transition considering interaction of $A_1$ and $B_1$ states can be written in the form

$$A_{ji} = \frac{<B_1, B_2 | V | A_1, B_2>}{E_f + E_c - E_{A_1} - E_{B_2} - i\gamma} <A_1, B_2 | V | c, f> \quad (4)$$

In the following, the matrix element of the Coulomb electron transfer between the levels of atoms 1 and 2 (numerator in (4)) will be briefly denoted by the symbol $W$, while the matrix element (3) by the symbol $V$. We get the intensities of the Auger transitions in different orders of scattering theory by calculating the terms of the series in the interaction.

The intensity of the Auger line as a function of the kinetic energy of the electron $I(\varepsilon)$ and its integral intensity $S$ (line power) in the lowest order are given by the equalities

$$I_0(\varepsilon) = \frac{V^2}{\pi} \frac{\gamma}{\varepsilon^2 + \gamma^2} ; \quad S_0 = \int d\varepsilon I_0 = V^2 . \quad (5)$$

The value $\varepsilon = E_f - (E_{A_1} + E_{B_2} - E_c)$ shows the deviation of the kinetic energy of the Auger electron $E_f$ from the value at maximum intensity.

Lifting holes from level $A_1$ to the neighboring atom level $B_1$ (Figure 1b) reduces the energy of the atomic residue by the amount $E_{21} = E_2 - E_1$ and gives a spectral line of the form

$$I_1(\varepsilon; E_{21}) = W^2 V^2 \frac{1}{\pi} \frac{1}{\varepsilon^2 + \gamma^2} \frac{\gamma}{(\varepsilon - E_{21})^2 + \gamma^2} ;$$
$$S_1 = \frac{2W^2 V^2}{E_{21}^2 + 4\gamma^2} ; \quad (6)$$

The virtual hole transition between levels 1↔2 (Figure 1c) produces additive

$$I_2(\varepsilon; E_{21}) = W^4 V^2 \frac{1}{\pi} \frac{\gamma}{(\varepsilon^2 + \gamma^2)^2} \frac{1}{(\varepsilon - E_{21})^2 + \gamma^2} ;$$
$$S_2 = \frac{W^4 V^2}{E_{21}^2 + 4\gamma^2} \{\frac{1}{2\gamma^2} + \frac{4}{E_{21}^2 + 4\gamma^2}\} . \quad (7)$$

Finally, the transfer of a hole from level $A_1$ to adjacent level $B_1$ of the accounting virtual transition $B_1 \leftrightarrow A_1$ generates the line

$$I_3(\varepsilon; E_{21}) = \frac{W^6 V^2}{\pi} \frac{1}{(\varepsilon^2 + \gamma^2)^2} \frac{\gamma}{((\varepsilon - E_{21})^2 + \gamma^2)^2} ,$$
$$S_3 = \frac{W^6 V^2}{(E_{21}^2 + 4\gamma^2)^2} (\frac{1}{\gamma^2} + \frac{16}{(E_{21}^2 + 4\gamma^2)}) . \quad (8)$$

**Figure 2** shows the contributions to the Auger spectra from the scattering channels considered above (equalities (5) - (8)) for level energy difference equal to their width $E_{21} = 2\gamma$. Thin solid line shows the intensity in the lowest order $I_0(\varepsilon)$. Intermittent lines depict the graphs of the functions $I_1(\varepsilon)$, $I_2(\varepsilon)$, $I_3(\varepsilon)$ (the first, the second and the third orders on the $W$ interaction of $A_1$ and $B_1$ states), and finally the bold solid line shows the sum of all four contributions. Allowance for the interaction of the levels of neighboring atoms leads to the significant increase in the Auger emission intensity in the case of $E_{21} = 2\gamma$. We add the intensities but not the amplitudes of the transitions, since a strong elastic scattering often leads to the randomization of the phases of the wave functions and, thus, to the weakening of interference effects.[13]

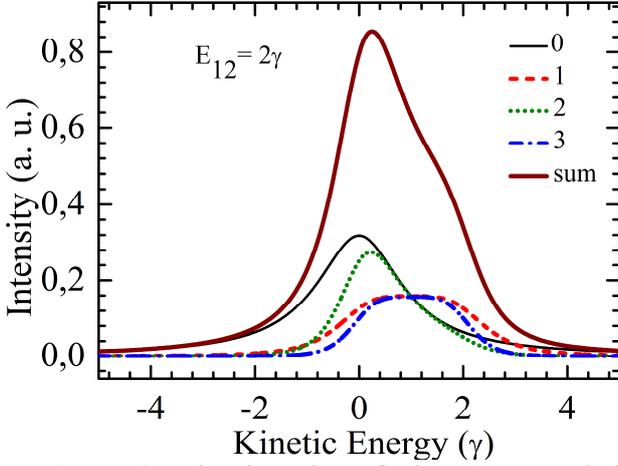

**Figure 2.** The intensity of the Auger emission $I(\varepsilon; E_{21})$ (the bold solid line) and the contributions of the channels (5) - (8) when the energy difference equal to the width of the levels $E_{12} = 2\gamma$.

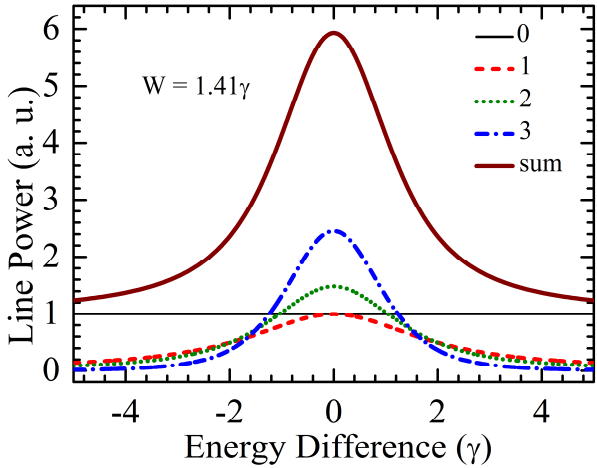

**Figure 3.** Integral intensity of the Auger line $S_0 - S_3$ as a function of the energy difference $E_{21}$ at $W = 1.41\gamma$: the power of the processes (5) - (8) and their sum (bold solid line).

**Figure 3** shows the dependence of the power of the processes $S_0 - S_3$ on the energy difference $E_{21}/\gamma$ with fixed value of the matrix element of the Coulomb transfer $W = 1.41\gamma$. The use of the finite series of perturbation theory for a small difference $E_{21}$ is not entirely justified, but even at $E_{21} = 2\gamma$ this series converges rapidly, and a two-three-fold increase in the Auger emission power due to the interaction of the internal levels of neighboring atoms is quite realistic. The enhancement effect decreases rapidly with the increasing energy difference and $E_{21} > 5\gamma$ becomes negligible. However, we must note that the above estimates were obtained in the adiabatic approach, although the production of holes triggers makes the process strongly nonequilibrium. We show below that the dynamic effect of a hole leads to the effective decrease in the difference $E_{21}$, increasing the probability of transitions between levels.

The presence of resonance (with a close energy) level on the neighboring atom substantially increases the probability of Auger transition. Contributions from the high-energy level deform the line in the direction of the increasing kinetic energy. Virtual excitations make the spectrum narrower. It should be considered that the final states represent a multiplet in real atoms, therefore usually there are not one but several lines with different energies.

## 3. Experimental Auger Spectra and Analysis

Single crystalline ingots of CuInSe2 were grown by the vertical Bridgman technique from near stoichiometric mixes of 99.999% purity elements in quartz ampoules sealed under vacuum without using any transport agent. Single crystalline bulk samples were taken from the middle parts of the ingots. The structure and the unit cell parameters have been characterized by the x-ray diffraction (XRD) technique as described in detail in Ref. [14]. The CuInSe2 samples were found to be single phase crystals in chalcopyrite structure.

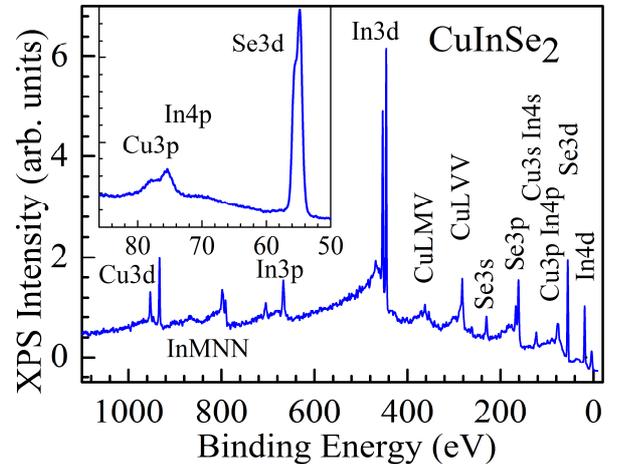

**Figure 4.** Survey spectrum of single-crystal CuInSe2, obtained at a photon energy of 1200 eV. The inset shows a fragment of the spectrum containing overlapping Cu3p and In4p lines (74-78 eV) and Se3d doublet (54.5-55.5 eV).

The XPS measurements were carried out using the synchrotron radiation at the Russian–German beamline of the BESSY II (Berlin). The analyzed samples were cleaved under high vacuum conditions in preparation chamber and then moved for XPS measurements without breaking vacuum. All the XPS spectra were measured under ultrahigh vacuum 2·10⁻¹⁰ mbar with the total-energy resolution set to 0.2 eV. The photonic beam was 20 μm in size. No noticeable charges of the sample surfaces by high-intensity X-ray beam has been observed during the measurements. All the measurements were carried out at room temperature.

**Figure 4** shows a survey spectrum of single crystal CuInSe2. The spectrum does not contain lines of extraneous elements. Oxygen is almost absent. Some carbon (282 eV) is deposited on the surface of the sample, but it does not affect the spectral regions of interest in the region of the Auger transitions. The inset shows a section with a binding energy of 50 - 86 eV. Se 3d doublet is a separate line, which indicates that the sample is single-phase. The overlapping Cu3p and In 4p levels are also visible.

**Figure 5** shows the Auger spectra of copper in CuInSe$_2$ and CuIn$_{0.9}$Ga$_{0.1}$Se$_2$ compounds, obtained at a photon energy of 1200 eV. The Auger spectrum of metallic copper[15] obtained on Mg Kα radiation of 1253.6 eV is also shown for comparison. The intra-atomic CuL$_3$VV Auger transitions (the kinetic energy maximum of 918 eV) and the triple Auger line formed by the CuL$_3$M$_{2,3}$V transition (the main maximum of 838 eV, the multiplet splitting due to the addition of the moments from the two holes 3p and 3d) are clearly visible on all the curves. At ~ 20 eV above the main lines, one can observe their replicas originating from the CuL$_2$ hole. At the same time, Auger spectra of compounds differ markedly from the spectrum of metallic copper. To understand the nature of the differences better, we first consider the electronic structure of the top-filled states.

seen from the valence band, then there is a peak from the In4d level with a binding energy of 17 eV and finally a Cu L$_3$VV Auger line in the range of 26-36 eV. Note that the latter energy shifts relative to its single-particle value by the magnitude of Hubbard repulsion of two valence holes on the copper atom. The Hubbard energy U measured in this way is 7 eV. A detailed evolution of the spectra in tuning the photon energy through the Cu L absorption edge was investigated in Ref. [16]. Here we note that the energy difference between the maxima of the valence band and the In4d level is approximately 15 eV and this level has great intensity with a small width.

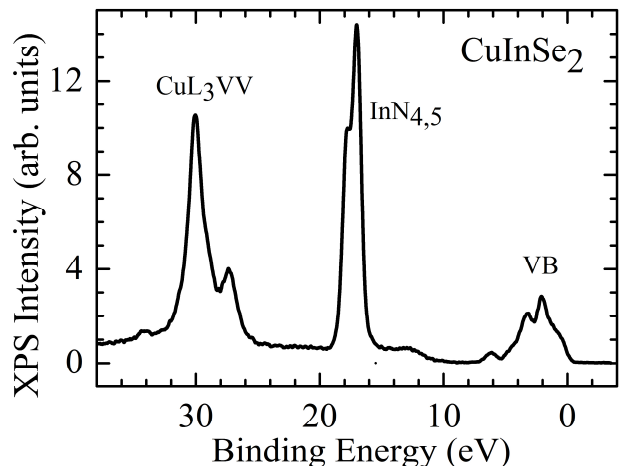

**Figure 6.** XPS spectrum of CuInSe$_2$ obtained at photon energies of 949.5 eV: a valence band and an In4d double peak with a binding energy of 17 eV. On the left is the Cu L$_3$VV Auger line, it is shifted by 7 eV with respect to its one-electron energy due to Hubbard repulsion of two 3d valence holes.

Figure 4 shows an additional signal beginning at 15 eV below the peak CuL$_3$VV in the compounds with indium which is absent in the pure copper spectrum. Its energy corresponds to the interatomic CuL$_3$InN$_{4,5}$V Auger transition. The same additional signal is visible in the compounds at 15 eV below the CuL$_3$M$_{2,3}$V Auger band. It is identified as the CuL$_3$M$_{2,3}$InN$_{4,5}$ interatomic Auger line in Figure 5.

The model developed in the previous section is applicable to the interatomic Auger transitions CuL$_3$M$_{2,3}$InN$_{4,5}$ and CuL$_3$InN$_{2,3}$N$_{4,5}$ directly. The intensity of the transitions is enhanced due to the resonant interaction of the levels of Cu3p and In4p, which are close in energy. Now look for the mechanism of amplification of the CuL$_3$InN$_{4,5}$V transition. In the next section, it will be shown that the transition to the final state with an In4d hole is likely due to electron shaking up in the Auger process.

**4. Electron Shaking up in Photoemission and Auger Transitions**

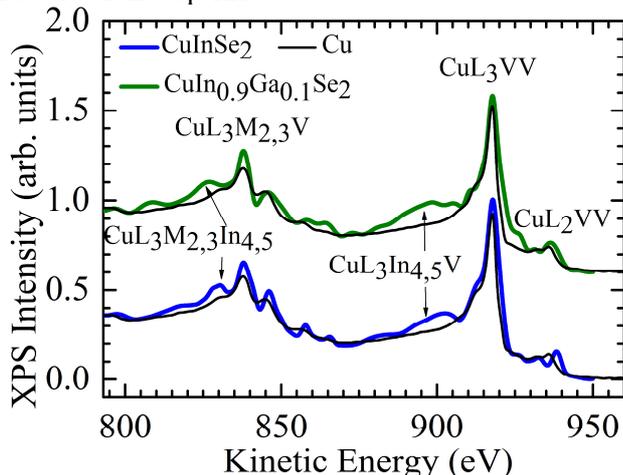

**Figure 5.** Intra-atomic CuL$_3$VV, CuL$_3$M$_{2,3}$V with interatomic CuL$_3$InN$_{4,5}$V, CuL$_3$M$_{2,3}$InN$_{4,5}$ Auger bands in compounds in CuInSe$_2$ and CuIn$_{0.9}$Ga$_{0.1}$Se$_2$ obtained at a photon energy of 1200 eV (thick lines) and pure copper spectrum[15] (Mg Kα radiation, thin line).

On the XPS spectrum of CuInSe2 (**Figure 6**) obtained at photon energies of 949.5 eV, a signal is The non-adiabatic process of hole creation gives rise to the appearance of the infinite number of electron-hole pairs near the Fermi energy in normal metals, which may emerge in spectra as singularity near the absorption and emission edges. This problem is dealt with in numerous publications reviewed in Ref.[17] We concluded that the hole effects are even stronger in the narrow band transition metals.[18,19] The semiconductor gap is not so important barrier to the formation of valence band electron-hole excitations in the dynamic field of the core-level photo-hole. Shaking is possible not only in valence band, but also from localized states.

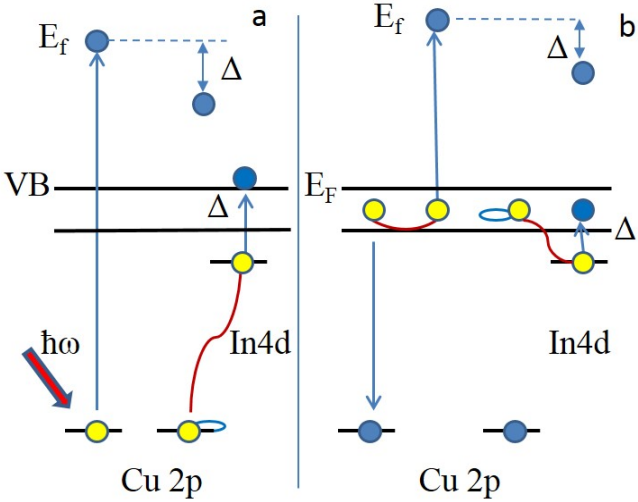

**Figure 7.** In4d electron shaking up ($\Delta$) accompanied direct Cu 2p photoemission (a) and Cu LVV Auger transition (b).

**Figure 7a** shows the creation of a hole on an indium atom upon the photoemission of electron from the 2p state on copper atom because of the second-order process. In the adiabatic approximation, the probability of the process is determined by the formula like (6)

$$I_1(\varepsilon;\Delta) = W^2 V^2 \frac{1}{\pi} \frac{1}{\varepsilon^2 + \gamma^2} \frac{\gamma}{(\varepsilon+\Delta)^2 + \gamma^2} \quad (9)$$

where $V$ is the matrix element of the Cu2p transition to the free state $f$ when the photon with energy $\hbar\omega$ is absorbed, the photoelectron energy $\varepsilon = E_f - \hbar\omega + E_{2p}$ measured relative to its value in the absence of losses, $\Delta = E_F - E_{4d} \approx 16$ eV is the excitation energy (in general, many-electron state), and $W$ stands for the Coulomb interaction of the suddenly produced Cu2p hole with In4d electron.

**Figure 7b** shows how the final state corresponding to the interatomic Auger transition is achieved in two stages. First, there is an intra-atomic Auger transition, and then the electron on neighbor indium atom is excited by the dynamic field of a hole. The process is again described by the equation (9) in which now

$\varepsilon = E_f - (E_V + E_V - E_{2p})$ is the addition to the energy of the intraatomic Auger transition with the matrix element $V$. The energy loss $\Delta$ varies within a few eV, beginning with the minimum value $\Delta = E_V - E_{4d}$, which is equal to the energy difference of the valence electron and the In4d electron.

Estimating the magnitude of the correction (9) to the main process (5), one can see that in the case of excitation energy $\Delta = 16$ eV and the level half-width $\gamma = 1$ eV, the second-order transition probability is very small. It should however be borne in mind that the formulas were obtained in the adiabatic approximation, while the x-ray excitation of the atom is strongly nonequilibrium process. A photo-hole appears in a very short time $\tau$, so it has a dynamic effect on surrounding electrons in a wide frequency range $\Delta\omega_h = 1/\tau$. The dynamic nature of the hole can be approximately considered by the introducing an alternating field of a hole, which significantly reduces difference $\Delta$. It is possible to replace it with an effective $\Delta_{eff} = E_F - E_{4d} - \hbar\omega_h$ in the denominator (9), which substantially increases the shakeup process. **Figure 8** demonstrates the shaking process in single crystal CuIn$_{0.5}$Ga$_{0.5}$Se$_2$. Indium substitution with isovalent gallium does not have a significant effect on the process under study since the Ga 3d state (binding energy 18.7 eV) is very close in type and energy to the corresponding 4d indium doublet (17.7 - 16.9 eV). Indeed, the Cu L$_{2,3}$ XPS spectrum in the compound shows the noticeable probability of shaking of In4d electrons. Satellites 16 eV below the main lines are visible in the compound and are absent in pure copper.

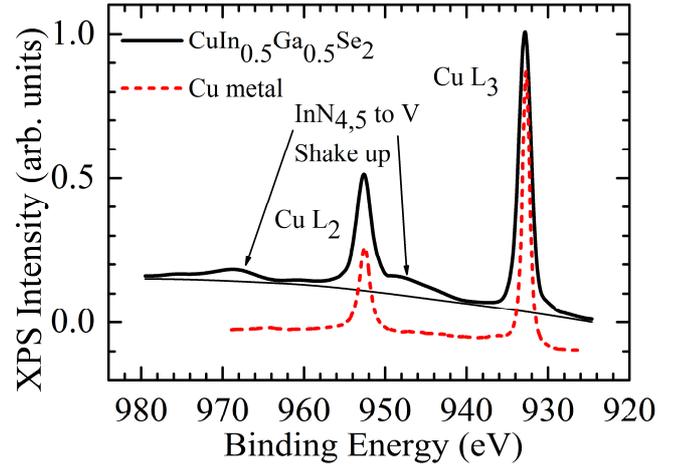

**Figure 8.** Cu L$_{2,3}$ XPS spectrum in the CuIn$_{0.5}$Ga$_{0.5}$Se$_2$ compound obtained at a photon energy of 1200 eV and pure copper spectrum[15] (dashed line). Satellites 16 eV below the main lines are visible in the compound and are absent in pure copper.

Shaking in the Auger process is even greater than in photoemission because there are not one, but two holes born there. The proposed mechanism explains the appearance of appreciable interatomic $CuL_3InN_{4,5}V$ Auger transitions in the compounds based on $CuInSe_2$.

## 5. Conclusion

The obtained experimental XPS spectra of compounds based on chalcopyrite CuInSe2 show intense interatomic $CuL_3M_{2,3}InN_{4,5}$ and $CuL_3InN_{4,5}V$ Auger transitions. The resonance enhancement of the interatomic Auger electron emission $CuL_3M_{2,3}InN_{4,5}$ is described by the multiple scattering theory with allowance for the energy proximity of Cu3p and In4p levels. The sudden appearance of holes in the photo and Auger emission creates a dynamic field with a wide frequency spectrum, which causes the shaking of electrons in the neighboring atoms. This process substantially increases the probability of the interatomic $CuL_3InN_{4,5}V$ Auger transition. In compounds with a narrow valence band (for example, 3d type) and a strong internal level with a not very high binding energy (In4d, 17 eV), a strong Coulomb interaction arises between electrons and holes on the neighboring atoms, which creates the favorable conditions for the appearance of intense interatomic transitions in the soft x-ray range.

### Acknowledgements

Our experiments were carried out at the Russian–German laboratory of the BESSY II synchrotron in Berlin. The study was supported by the Russian Science Foundation (project no. 17-12-01500).